\documentclass[epj]{svjour}
\usepackage{epsfig}
\usepackage{graphics}
\usepackage{graphicx}
\newcommand{\boldsymbol}{\bf}
\newcommand{\bs}{\bf}
\newcommand{\ar}{\arrowvert}

\newcommand{\be}{\begin{equation}}
\newcommand{\ee}{\end{equation}}
\newcommand{\ba}{\begin{eqnarray}}
\newcommand{\ea}{\end{eqnarray}}

\begin{document}
\title{QCD Coulomb Gauge Approach to Exotic Hadrons}
\author{Stephen R. Cotanch, Ignacio J. General and  Ping Wang
}                     
%
%
\institute{Department of Physics, North Carolina State University, Raleigh NC 27695-8202 USA}
\date{Received: date / Revised version: date}
%
\abstract{ The Coulomb gauge Hamiltonian model is used to
calculate masses for selected $J^{PC}$ states consisting of exotic
combinations of quarks and gluons: $ggg$ glueballs (oddballs), $q
\bar{q} g$ hybrid mesons and $q \bar{q} q \bar{q}$ tetraquark
systems.  An odderon Regge trajectory is computed for the $J^{--}$
glueballs with intercept much smaller than the pomeron, explaining
its nonobservation.  The lowest $1^{-+}$ hybrid meson mass is
found to be just above 2.2 GeV while the lightest tetraquark state
mass with these exotic quantum numbers is predicted around 1.4 GeV
consistent with the observed $\pi(1400)$.
} 
\titlerunning{QCD Coulomb Gauge Approach}
\maketitle
\section{Introduction}
\label{intro}
Establishing the existence of exotic hadrons (non $q \bar{q}$ or $qqq$ structure) is of paramount importance and remains one of the key unsolved problems in hadronic physics.  In particular, it is expected from general QCD principles  that nonconventional color singlet states of
gluons and quarks should exist such as glueballs ($gg$ and $ggg$), hybrid mesons ($q \bar{q} g$) and multiquark states ($q \bar{q} q \bar{q}$
and $qqqq \bar{q}$).  The present work addresses the structure of these hadrons and provides new information  to assist experimental searches.
\section{Coulomb Gauge Hamiltonian Model}
\label{sec:1}
In a series of publications \cite{Szczepaniak:1995cw,LC1,LChybrid,LC2,Llanes-Estrada:2000jw,LCSS,LBC,gcl} a realistic model for hadron structure has been developed and  applied to the quark and gluon sectors.
This field
theoretical, relativistic many-body approach utilizes an
effective QCD Hamiltonian, $H_{\rm eff}$, formulated in the
Coulomb gauge.  It properly incorporates
chiral symmetry using
standard bare current  quark masses  but dynamically generates a constituent mass and
spontaneous chiral symmetry breaking \cite{LC1}. Through approximate many-body diagonalizations it  successfully describes
the meson spectrum \cite{LC2,LCSS} and is   consistent
\cite{LBC} with lattice  glueball  predictions.  It also
yields a good description of the vacuum properties (quark and
gluon condensates) within a minimal two parameter theory.

The effective Hamiltonian, an approximation to the
exact Coulomb gauge QCD Hamiltonian, is
\begin{eqnarray}
H_{\rm eff} &=& H_q + H_g +H_{qg} + H_{C}   \\
H_q &=& \int d{\bf x} \Psi^\dagger ({\bs x}) [ -i {\mbox{\boldmath$\alpha$\unboldmath}} \cdot {\mbox{\boldmath$\nabla$\unboldmath}}
+  \beta m] \Psi ({\bs x})   \\
H_g &=& \frac{1}{2} \int d {\bs x}\left[ {\bf \Pi}^a({\bs x})\cdot {\bf
\Pi}^a({\bs x}) +{\bf B}^a({\bs x})\cdot{\bf B}^a({\bs x}) \right] \\
H_{qg} &=&  g \int d {\bs x} \; {\bf J}^a ({\bs x})
\cdot {\bf A}^a({\bs x}) \\
H_C &=& -\frac{1}{2} \int d{\bs x} d{\bs y} \rho^a ({\bs x}) \hat{V}(\ar {\bs x}-{\bs y}
\ar ) \rho^a ({\bs y})   \ ,
 \label{model}
\end{eqnarray}
with $g$  the QCD coupling, $\Psi$  the quark field, $m$ the
current quark mass, ${\bf A}^a$  the gluon fields
satisfying the transverse gauge condition,
$\mbox{\boldmath$\nabla$\unboldmath}$ $\cdot$ ${\bf A}^a = 0$, $a
= 1, 2, ... 8$, ${\bf \Pi}^a $  the conjugate fields and ${\bf
B}^a$  the non-abelian magnetic fields
\begin{eqnarray}
{\bf B}^a = \nabla \times {\bf A}^a + \frac{1}{2} g f^{abc} {\bf A}^b \times {\bf A}^c \ .
\end{eqnarray}
The color densities, $\rho^a({\bs x})$, and currents, ${\bf J}^a$, are
\begin{eqnarray}
\rho^a({\bs x}) &=& \Psi^\dagger({\bs x}) T^a\Psi({\bs x}) +f^{abc}{\bf
A}^b({\bs x})\cdot{\bf \Pi}^c({\bs x}) \\
{\bf J}^a &=& \Psi^\dagger ({\bs x}) \mbox{\boldmath$\alpha$\unboldmath}T^a \Psi ({\bs x})
\ ,
\end{eqnarray}
with
$T^a = \frac{\lambda^a}{2}$ and $f^{abc}$   the
$SU_3$ color matrices and structure constants, respectively.
Confinement is described by a Cornell type potential,
\ba
\label{2}
\hat{V} (r = |{\bs x} - {\bs y}|) &=& {\hat V}_C (r) + {\hat V}_L (r) \\
    \hat{V}_C(r) &=& -\frac{\alpha_s}{r} \\
 \hat {V}_L(r)   &=& \sigma r ,
\ea
with previously determined string tension,  $\sigma=0.135$ GeV$^{2}$, and
$\alpha_s=\frac{g^2}{4\pi}=0.4$.  Below we denote
the Fourier transform of $\hat V$  by $V$.

The bare  fields have the Fock operator expansions
(quark spinors $u, v$, helicity, $\lambda = \pm 1$, and
color vectors $\hat{\bf{\epsilon}}_{{\cal C }= 1,2,3}$)
\begin{eqnarray}
\label{colorfields1}
 \Psi({\boldsymbol{x}}) &=&\int \!\! \frac{d
    {\boldsymbol{k}}}{(2\pi)^3} \Psi_{{\cal C}} ({\boldsymbol{k}})  e^{i \boldsymbol{k} \cdot \boldsymbol{x}} \hat{\boldsymbol{\epsilon}}_{\cal C}  \\
\Psi_{ {\cal C}} (\boldsymbol{k})   & = & {u}_{\lambda} ({\boldsymbol{k}}) b_{\lambda {\cal C}}({\boldsymbol{k}  )}+ {v}_{\lambda} (-{\boldsymbol{k}})
    d^\dag_{\lambda {\cal C}}(\boldsymbol{-k)}   \\
{\bf A}^a({\bs{x}}) &=&  \int
\frac{d{\bs{k}}}{(2\pi)^3}
\frac{1}{\sqrt{2k}}[{\bf a}^a({\bs{k}}) + {\bf a}^{a\dag}(-{\bs{k}})]
e^{i{\bs{k}}\cdot
{\bs {x}}}  \ \ \
\\
{\bf \Pi}^a({\bs{x}}) &=& \hspace{-.15cm}-i \int \!\!\frac{d{\bs{k}}}{(2\pi)^3}
\sqrt{\frac{k}{2}}
[{\bf a}^a({\bs{k}})-{\bf a}^{a\dag}(-{\bs{k}})]e^{i{\bs{k}}\cdot
{\bs{x}}}  \!,
\end{eqnarray}
with quark, anti-quark and gluon Fock operators
$b_{\lambda {\cal C}}(\boldsymbol{k)}$, $d_{\lambda {\cal
C}}(\boldsymbol{-k)} $ and  $a_{\mu}^a({\bs{k}})$ ($\mu = 0, \pm
1$), respectively.
The Coulomb gauge  condition,
${\bs k}\cdot {\bf
a}^a ({\bs k}) =  (-1)^\mu k_{\mu} a_{-\mu} ^a ({\bs k}) =0$,
produces  transverse commutation relations,
\begin{equation}
[a^a_{\mu}({\bs k}),a^{b \dagger}_{\mu'}({\bs k}')]=
(2\pi)^3 \delta_{ab} \delta^3({\bs k}-{\bs k}')D_{{\mu} {\mu'}}({\bs k})  \ ,
\end{equation}
with
\begin{equation}
D_{{\mu} {\mu'}}({\bs k}) =
\delta_{{\mu}{\mu'}}- (-1)^{\mu}\frac{k_{\mu} k_{-\mu'}}{k^2}  \  .
\end{equation}

The ground state (model vacuum) is generated using the Bardeen-Cooper-Schriffer (BCS) method,
entailing rotated field operators
(Bogoliubov-Valatin transformation),
\begin{eqnarray} \label{eq:operator rotations}
    B_{\lambda {\cal C}}(\boldsymbol{k)} &=& \cos\frac{\theta_k}{2}
    b_{\lambda {\cal C}}({\boldsymbol{k})}  - \lambda \sin\frac{\theta_k}{2}
    d^\dag_{\lambda {\cal C}}(\boldsymbol{-k)}   \nonumber  \\ \nonumber
    D_{\lambda {\cal C}}(\boldsymbol{-k)}&=& \cos\frac{\theta_k}{2}
    d_{\lambda {\cal C}}({\boldsymbol{-k})}  + \lambda \sin\frac{\theta_k}{2}
    b^\dag_{\lambda {\cal C}}(\boldsymbol{k)}  \\
    {\boldsymbol \alpha}^a(\boldsymbol{k}) &=& \cosh \Theta_k {\bf a}^a({\boldsymbol{k}}) + \sinh \Theta_k
    {{\bf a}^a}^\dag(-\boldsymbol{k}) \ ,
\end{eqnarray}
producing the dressed,
quasi-particle operators ${\boldsymbol \alpha}^a$, $B_{\lambda {\cal C}}$ and $D_{\lambda {\cal C}}$, respectively.
The quasi-particle (BCS) vacuum, determined by $B_{\lambda {\cal C}} |\Omega
\rangle = D_{\lambda {\cal C}} |\Omega\rangle = \alpha^a_\mu |\Omega\rangle = 0$,  is
built on the bare parton one, $b_{\lambda {\cal C}} |0\rangle = d_{\lambda {\cal C}} |0\rangle = a^a_\mu
|0\rangle = 0$,
\begin{equation}
    |\Omega_{quark}\rangle =e^{- \int \!\!
    \frac{d\boldsymbol{k}}{(2\pi)^3}\lambda
    \tan\frac{\theta_k}{2} b^\dag_{\lambda {\cal C}}({\boldsymbol{k}})
    d^\dag_{\lambda {\cal C}} (-\boldsymbol{k})} |0\rangle   \nonumber
\end{equation}
\begin{equation}
    |\Omega_{gluon}\rangle = e^ {- \!\! \int \!\!
    \frac{d\boldsymbol{k} }{(2\pi)^3} \frac{1}{2}\tanh\Theta_k D_{\mu \mu'}({\bs k})
    a_{\mu}^{a\dag} ({\boldsymbol{k}}) a_{\mu'}^{a\dag} (-\boldsymbol{k})} |0\rangle \ . \nonumber
\end{equation}
The composite BCS vacuum, $|\Omega \rangle = |\Omega_{quark}
\rangle \otimes |\Omega_{gluon} \rangle $,  contains
quark and gluon condensates (correlated $q\bar{q}$ and $gg$ Cooper pairs).
 A variational minimization of the vacuum
expectation value of the Hamiltonian, $\delta \langle\Omega|H_{\rm eff}|
\Omega\rangle = 0$,
yields the constituent quark and gluon  gap equations
\ba
    k s_k -  m c_k =&& \frac{2}{3}\int \!\!
    \frac{d\boldsymbol{q}}{(2\pi)^3}
    ( s_k c_q x - s_q c_k)
V(|\boldsymbol{k-q}|)
\ea
\ba
\label{ggapeq}
    \omega_k^2 &=& k^2-\frac{3}{4}\int \!\!
    \frac{d\boldsymbol{q}}{(2\pi)^3}
V(|{\boldsymbol{k-q}}|)[1+x^2]
    \bigg( \frac{\omega_q^2-\omega_k^2}{\omega_q} \bigg) \nonumber \\
    &+&  \frac{3}{4} \; g^2\int \!\!
    \frac{d\boldsymbol{q}}{(2\pi)^3} \frac{1 - x^2}{\omega_q} \ ,
\ea
with $s_k = sin \phi_k$, $c_k = cos \phi_k$ and $x = {\bs k
}\cdot {\bs q}$.
Here $\phi_k = \phi (k)$ is  the quark gap angle
related to the
BCS angle $\theta_k$ by, $tan(\phi_k  - \theta_k ) = m/k$, and
$\omega_k = k e^{-2\Theta_k}$ is the effective  gluon self energy.
The last term in Eq. (\ref{ggapeq}) is due to
the non-abelian component of the gluon kinetic energy.
The quark gap
equation is UV finite for the linear potential since
$V_L(|{\boldsymbol{p}}|) = - 8 \pi \sigma /p^{4}$, but
not for the Coulomb potential
$V_C(|{\boldsymbol{p}}|) = - 4 \pi \alpha_s /p^{2}$.   The gluon gap equation has both logarithmical and quadratical UV
divergences  and an integration cutoff, $\Lambda
= 4$ GeV, determined in previous studies is used in both equations.

\section{Applications}
Predictions for the low-lying spectra of glueballs, hybrid mesons and tetraquark systems are now presented and discussed.  Since these hadrons consist of 3 or more constituents, the masses for selected $J^{PC}$ states are computed variationally
\ba
\label{varyeq}
M_{J^{PC}}& =& \frac{ \langle \Psi^{JPC} | H_{\rm eff} | \Psi^{JPC} \rangle} {\langle\Psi^{JPC}|\Psi^{JPC}\rangle} \ .
\ea
The variational approximation has been comprehensively tested in two body systems by comparison with exact diagonalization and found to be  accurate to  a few percent.
\subsection{Glueballs}
\label{sec:2}
\begin{figure} [b]
\vspace{-7cm}
\hspace{3cm}
\includegraphics[width=3.75cm]{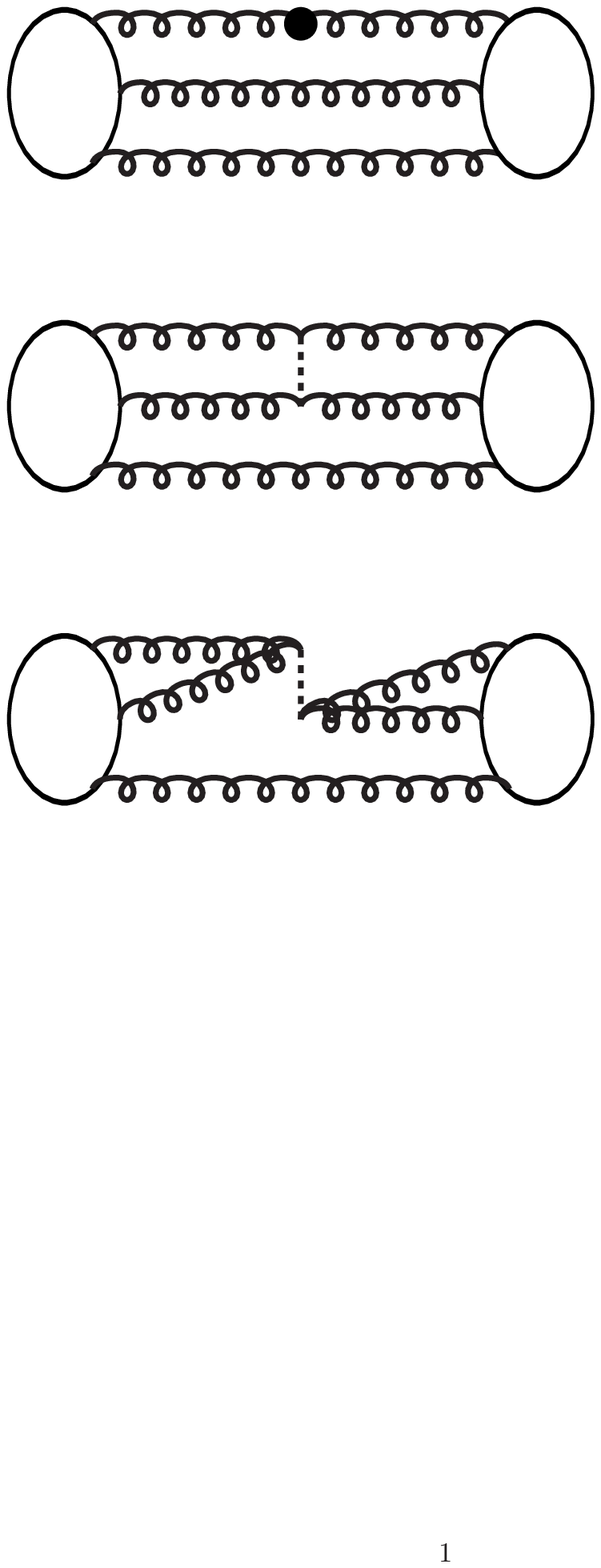}
\caption{\label{TDA3g} Glueball diagrams for $\langle \Psi^{JPC}_{ggg} |H_{\rm eff}| \Psi^{JPC}_{ggg} \rangle$.}
\end{figure}

Previous work~\cite{Llanes-Estrada:2000jw} has investigated  $gg$ glueballs which only have $C = 1$.
For $C =  -1$ glueballs (oddballs),  Fock states with at least 3 gluons are necessary and
the  variational wavefunction  is (${\bf q}_{i = 1,2,3}$  are the $cm$ gluon momenta)
\begin{eqnarray}
\lefteqn{\hspace{.5cm}\arrowvert \Psi^{JPC}_{ggg} \rangle = \int d{\bf q}_1 d{\bf q}_2
d{\bf q}_3
\delta({\bf q}_1 + {\bf q}_2 + {\bf q}_3 )} \\
& &
\hspace{-.3cm}
\Phi^{JPC}_{\mu_1 \mu_2 \mu_3}({\bf q}_1,{\bf q}_2,{\bf q}_3)
C^{abc}
\alpha^{a\dagger}_{\mu_1}({\bf q}_1)
\alpha^{b\dagger}_{\mu_2}({\bf q}_2) \alpha^{c\dagger}_{\mu_3}({\bf q}_3)
\arrowvert \Omega_{gluon} \rangle \ , \nonumber
\end{eqnarray}
with
color tensor $C^{abc}$   either totally
antisymmetric $f^{abc}$ (for $C = 1$) or symmetric
$d^{abc}$ (for $C = -1$).
Boson statistics thus requires the $C = -1$ oddballs to have a
symmetric space-spin wavefunction.
Using eq. (\ref{varyeq}) and a two-parameter variational radial wavefunction,
the $J^{--}$ oddball states have been calculated.  Only the Abelian component of the magnetic fields
ae retained and the hyperfine interaction, $H_{qg}$, is suppressed. There are three
terms contributing to the mass expectation value which are depicted in Fig.  \ref{TDA3g}
which correspond to the gluon self-energy (top), gluon-gluon scattering (middle) and
annihilation (bottom).

The nine-dimensional   variational calculation
was performed using the Monte Carlo method with  the
adaptive sampling algorithm VEGAS~\cite{Vegas} and numerical
convergence required between  $10^5$ and $10^6$ samples.
The oddball mass predictions are compared in Table~\ref{statetable}  to available

\newpage
\noindent
lattice gauge results~\cite{mp,Meyer:2004jc} and a Wilson-loop inspired model~\cite{ks}.
A study
of the glueball mass sensitivity to both statistical and variational
uncertainties yielded error bars at the few per cent level.

\begin{table}[t]
\caption{\label{statetable} Glueball  quantum numbers and  masses
in MeV. Error in  $H_{\rm eff}$ (from Monte Carlo only) is
100 MeV or less, the quoted
lattice errors are typically 200-300
MeV.
}
\begin{tabular}{|c|cccc|}
\hline
 Model &  $1^{--}$ &  $3^{--}$ & $5^{--}$ & $7^{--}$
\\ \hline
Coulomb gauge $H_{\rm{eff}}$&   3950  & 4150 & 5050 & 5900  \\
lattice~\cite{mp}&  3850   &4130 & &  \\
lattice~\cite{Meyer:2004jc}& 3100  &4150 & &  \\
Wilson-loop~\cite{ks}& 3490  & 4030 & & \\
\hline
\end{tabular}
\end{table}

\begin{figure}[h]
\vspace{-.65cm}
\psfig{figure=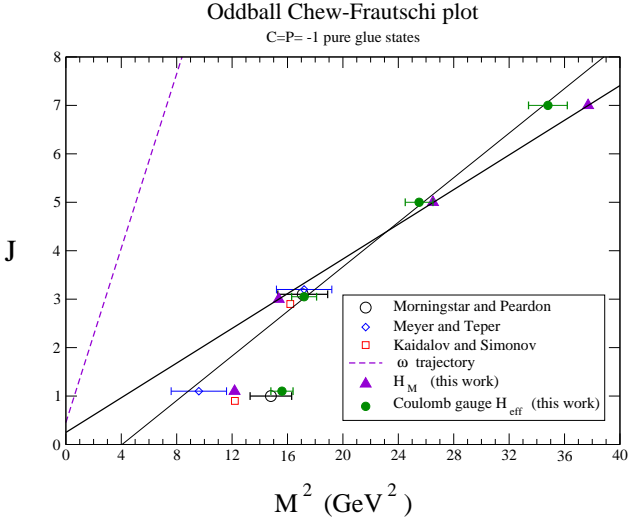,width=2.9in,angle=-90}
\caption{\label{odderonfig} Odderon trajectories
from constituent gluon models and lattice compared to the
$\omega$ meson Regge trajectory.}
\end{figure}

Figure \ref{odderonfig}
displays predicted oddball Regge trajectories from the alternative
approaches.
Lattice results are depicted by  open
circles~\cite{mp} and diamonds~\cite{Meyer:2004jc}.   Constituent gluon predictions
are represented by boxes, solid  triangles and solid circles and correspond
to   a Wilson-loop inspired potential model~\cite{ks},  a simpler harmonic oscillator
calculation~\cite{LBC} labeled $H_{\rm M}$ and
the  $H_{\rm eff}$ approach, respectively.
The   odderon trajectories for the latter two models are represented by the solid
lines,  $\alpha_O^{\rm M} = 0.18t + 0.25$ and $\alpha_O^{\rm eff} = 0.23 t - 0.88$,
 which provide an overall
theoretical uncertainty. The
much steeper dashed line is the
$\omega$ trajectory.

Three key results follow.  First,   the
predicted   odderon  has slope
similar
to the pomeron but   intercept clearly lower than
the $\omega$ value.  Second, the odderon starts
with the $3^{--}$ state and  not the $1^{--}$ which is
on a daughter trajectory.  Note that there are no lattice
$5^{--}$  glueball predictions which are necessary to confirm this point
and we strongly recommend that
future studies calculate  higher $J^{--}$ states.
Third, all approaches agree that the
$3^{--}$ mass is near  4 GeV.

\subsection{Hybrid mesons}
We denote the momenta of the
dressed quark, anti-quark and gluon by ${\bs q}$, $ { \bar{\bs q}}$
and ${\boldsymbol g}$, respectively,
and work in the
hybrid $cm$ system. The color structure for a $q\bar{q}g$ hybrid is given by
$SU_c(3)$ algebra,
   $ (3 \otimes \overline{3}) \otimes 8 = ( 8\otimes 8) \oplus (8\otimes 1)
    = 27 \oplus 10 \oplus 10 \oplus 8 \oplus 8
    \oplus 8 \oplus 1$.
Hence for an overall color singlet  the
quarks must  be in an octet state like the gluon
which leads to a repulsive $q \bar{q}$ interaction, confirmed by lattice at short range, that
raises the mass of the hybrid meson. The hybrid wavefunction has
the general form
  \ba
    |\Psi^{JPC}_{q\bar{q}g} \rangle = \int \!\!  {d\boldsymbol{q}
   d\boldsymbol{\bar q} d\boldsymbol{g} } \; \delta ({\bs q + \bar q + g}) \; {\Phi^{JPC}_{\lambda {\bar \lambda} \mu}(\boldsymbol{q},\boldsymbol{\bar q}, \boldsymbol{g})}  \nonumber
    \\ T_{{\cal C} { \bar {\cal  C}}}^a B^ \dag_{\lambda {\cal C} }({\boldsymbol{q}})
    D^ \dag_{\bar {\lambda} { \bar {\cal  C}} }(\overline{\boldsymbol{q}}) \alpha^{a\dag}_{\mu}({ \boldsymbol g}) |\Omega \rangle  \ .
  \ea

\begin{figure} [h]
    \hspace{.2cm}
   \includegraphics[width=.45\textwidth]{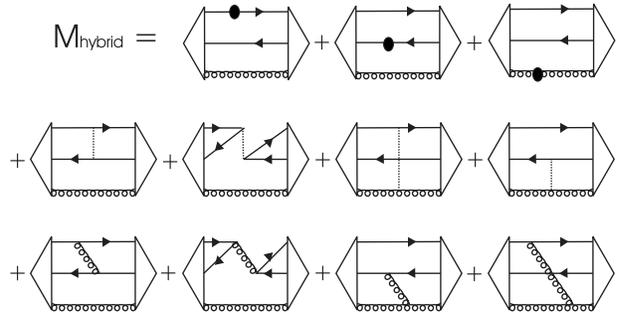}
       \vspace{.5cm}
    \caption{Hybrid meson diagrams for $\langle \Psi^{JPC}_{q\bar{q}g} |H_{\rm eff}| \Psi^{JPC}_{q\bar{q}g} \rangle$.}
    \label{hybriddiagrams}
\end{figure}

We have extended our previous hybrid study \cite{LChybrid} by including
the $H_{qg}$ Hamiltonian term  containing
the ${\bf J}^a \cdot {\bf A}^a$ operators.
Following \cite{LCSS},
an effective quark hyperfine
interaction with a ${\bf J}^a \cdot {\bf J}^a$ form is
obtained using
perturbation theory to second order in $g$ and integrating over
the gluonic degrees of freedom. This contribution to the hybrid mass is
represented by the $q \bar{q}$ gluon exchange Feynman diagrams in Fig. \ref{hybriddiagrams}
(first two in the bottom row).  The non-abelian magnetic field terms are also included
and entail triple-gluon vertices (last two diagrams in Fig. \ref{hybriddiagrams}).  The
remaining diagrams represent the self-energy,  scattering and quark annihilation mass contributions.

The  hyperfine interaction from the $H_{qg}$ term is
\begin{equation} \label{JJdiagram}
V_{T} =\frac{1}{2}\int \!\!\! \int \!\!
d{\boldsymbol{x}}{d\boldsymbol{y}}
    J_i^a({\boldsymbol{x}}) \hat{U}_{i j}({\boldsymbol{x,y}}) J_j^a(\boldsymbol{y}),
\end{equation}
with kernel reflecting the transverse gauge
\begin{equation}
    \hat{U}_{i j}({\boldsymbol{x,y}}) = \bigg(\delta_{i j}- \frac{\nabla_i
    \nabla_j}{\nabla^2}\bigg)_{\boldsymbol{x}}\hat{U}(|\boldsymbol{x-y}|).
\end{equation}
The potential $\hat{U}$ is a modified Yukawa   with
dynamical mass, $m_g = 600$ MeV, for the exchanged gluon as
explained in \cite{LCSS}. Its momentum space representation is
\begin{equation} \label{Adam's potential}
    { U}(p) = \left\{ \begin{array}{ll}
        -\frac{8.04}{p^2}
    \frac{\ln^{-0.62}(\frac{p^2}{m_g^2}+0.82)}{\ln^{0.8}(\frac{p^2}{m_g^2}+1.41)}& \textrm{ $p>m_g$}\\
        -\frac{24.50}{p^2+m_g^2} & \textrm{$p<m_g$}
        \end{array} \right. \ .
\end{equation}
The quark hyperfine interaction also generates additional terms in the quark gap equation
\cite{LCSS}.

The 12 dimensional integrals were calculated
using the Monte Carlo method and repetitively evaluated
with an increasing number of points until a weight-averaged result
converged, typically involving about 50 million samples. The hybrid mass error introduced by this procedure  is
about $\pm$ 50 MeV. For each $J^{PC}$ hybrid state we optimized
the two variational parameters.

\begin{figure} [h]
\includegraphics[width=0.45\textwidth]{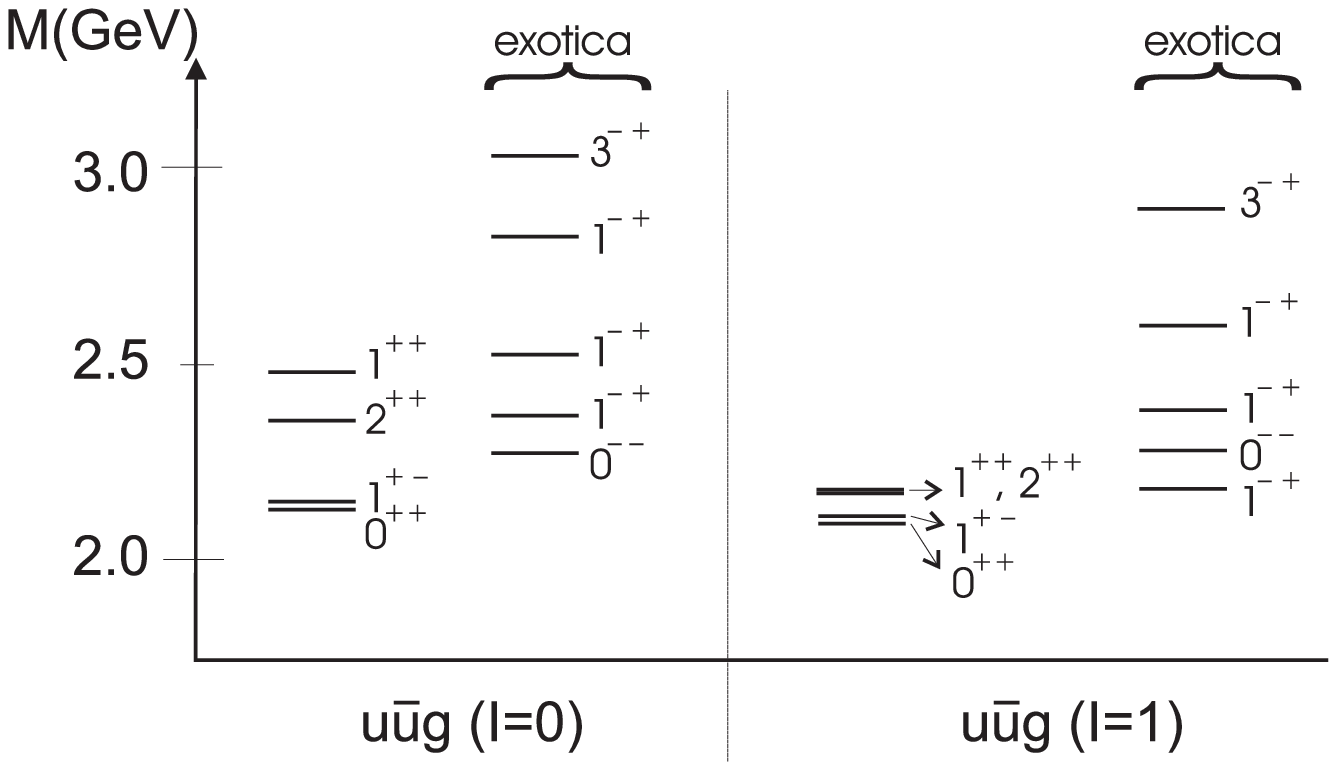}
  \caption{Low lying $u \overline{u}g$ spectra.}
  \label{resultsUU}
\end{figure}

\begin{figure} [h]
  \includegraphics[width=.45\textwidth]{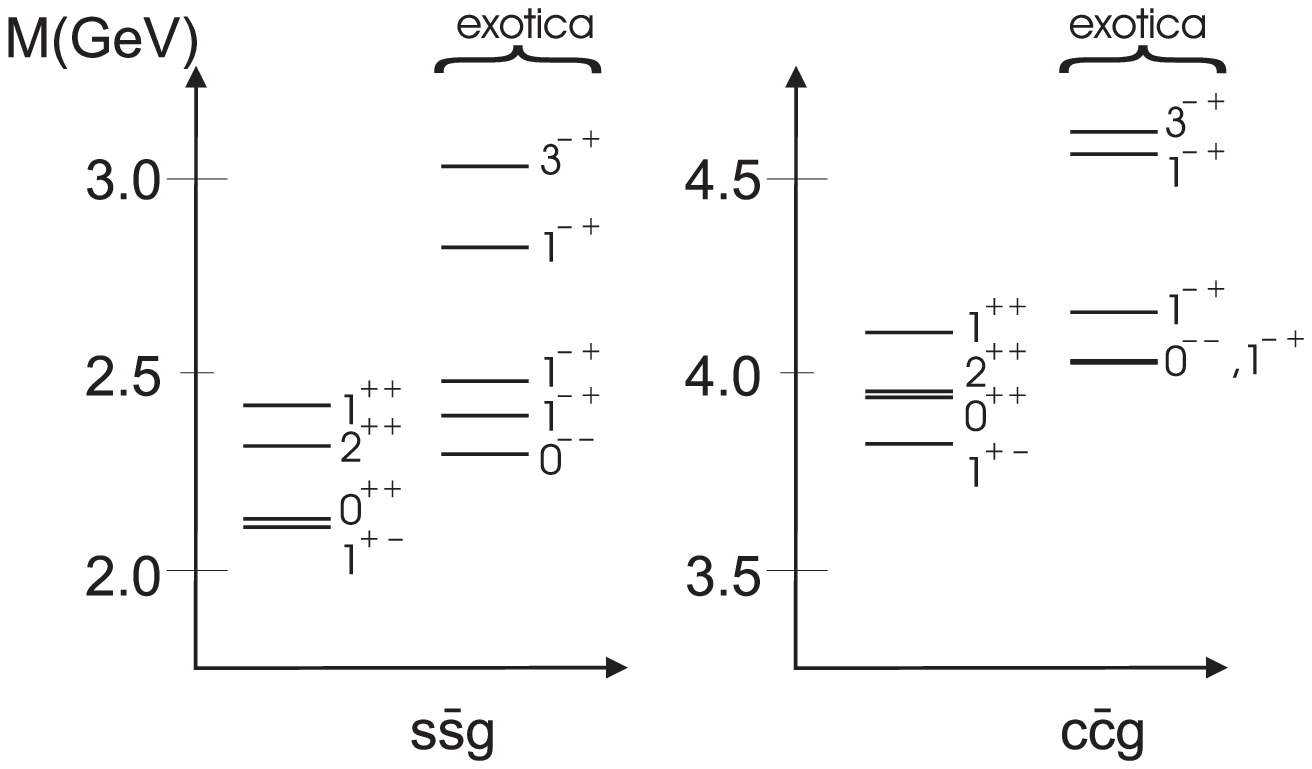}
   \caption{Low lying $s \overline{s}g$ and $c \overline{c}g$ spectra.}
   \label{resultsSS-CC}
\end{figure}

Using standard current quark masses, $m_u = m_d = 5$ MeV, $m_s = 80 $ MeV and $m_c = 1000 $ MeV, the predicted low-lying  mass spectra for light and heavy hybrid mesons
are presented in Figs. \ref{resultsUU} and \ref{resultsSS-CC}, respectively.
Note that quark annihilation interactions increase the hybrid mass and this introduces  isospin
splitting since it only contributes in the $I_{q \bar {q}}$ channel.
More importantly, all hybrid masses, especially the lightest exotic $1^{-+}$ state, are clearly above 2 GeV.
This is consistent with lattice  \cite{Bernard1,Bernard2,Lacock,Hedditch,Luo,Liu,Griffiths,Perantonis}  and Flux Tube model \cite{Barnes,Close,Katja}  results summarized in Table \ref{table:comparison-u}.
These composite predictions strongly suggest that observed $1^{-+}$ $\pi(1600)$, and more clearly $\pi(1400)$, are not hybrid meson states.

The charmed $c \bar c g$ hybrid spectrum has a slightly different
level order compared to the strange $s \bar s g$ and $u \bar u g$
spectra  due to the hyperfine interaction. The predicted strange
and charmed exotic $1^{-+}$ states are also in reasonable
agreement with both lattice and Flux Tube results.

\begin{table} [t]
\caption{Published predicted exotic  $1^{-+}$  masses, in GeV, for light, strange and charmed hybrid mesons.}
  \label{table:comparison-u}
  \begin{tabular}{|c|c|c|c|}
  \hline
Model &    $u/d$ hybrid & $s$ hybrid & $c$ hybrid   \\
    \hline
Lattice QCD [13-20] &   1.7 - 2.1 & 1.9 &  4.2 - 4.4   \\
Flux Tube  \cite{Barnes,Close,Katja}  &    1.8 - 2.1 & 2.1 - 2.3  & 4.1 - 4.5 \\
\hline
  \end{tabular}
\end{table}
%
%
%

\subsection{Tetraquark systems}
\begin{figure} [h]
\hspace{.05cm}
    \includegraphics[width=.45\textwidth]{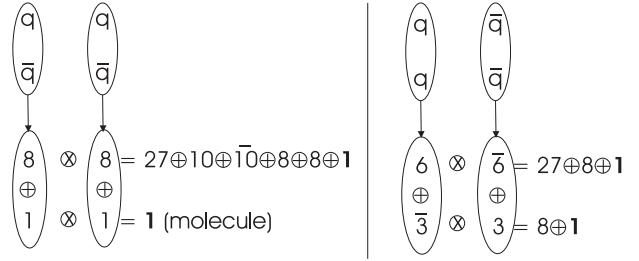}
    \caption{Color singlets  via four different representations.}
    \label{colorschemes}
\end{figure}

This is the first four-body application using this approach.
The $SU_c(3)$ color algebra for four quarks produces 81 color states,
$3\otimes\overline3\otimes3\otimes\overline3=27\oplus10\oplus
\overline{10}\oplus8\oplus8\oplus8\oplus8\oplus1\oplus1$, of which two are color singlets that
can be obtained in four different ways, depending on the
intermediate color coupling: singlet scheme (molecule),
octet scheme, and two diquark schemes involving the triplet  and the sextet representations (see
Fig. \ref{colorschemes}).

Working in the  $cm$ system and denoting
the momenta of the quarks by ${\bf q_1}$ and ${\bf q_3}$, and
those of the anti-quarks by ${\bf q_2}$ and ${\bf q_4}$,
the tetraquark wavefunction is
  \ba
\lefteqn{    |\Psi^{JPC}_{4q}\rangle =  \int \!\!
    {d\boldsymbol{q}_1}{d\boldsymbol{q}_2}
    {d\boldsymbol{q}_3} {d\boldsymbol{q}_4}
    \delta({\bf q}_1 + {\bf q}_2 + {\bf q}_3 + {\bf q}_4 )} \nonumber  \\
& &
  \hspace{2.5cm}  \Phi^{JPC}_{\lambda_1 \lambda_2
    \lambda_3 \lambda_4}({\boldsymbol{q}_1,\boldsymbol{q}_2,\boldsymbol{q}_3, \boldsymbol{q}_4})
    R^{{\cal C}_1{\cal C}_2}_{{\cal C}_3{\cal C}_4}  \\
    & &
     B^{\dag}_{\lambda_1{\cal C}_1}({\boldsymbol{q}_1})
   { D^{\dag}_{\lambda_2{\cal C}_2}(\boldsymbol{q}_2)}
   { B^{\dag}_{\lambda_3{\cal C}_3}(\boldsymbol{q}_3)}
    {D^{\dag}_{\lambda_4{\cal C}_4}(\boldsymbol{q}_4)}|\Omega_{quark} \rangle \ , \nonumber
  \ea
  where the color elements $R^{{\cal C}_1{\cal C}_2}_{{\cal C}_3{\cal C}_4}$
depend on the specific color scheme chosen.

\begin{figure} [h]
    \hspace{.04cm}
    \includegraphics[width=.42\textwidth]{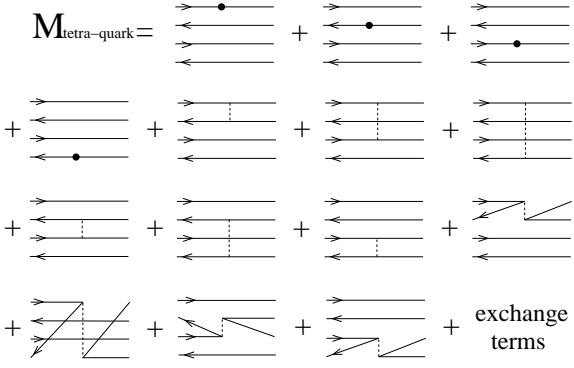}
     \hspace{1cm}
    \caption{Tetraquark diagrams for $\langle\Psi^{JPC}_{4q}|H_{\rm eff}|\Psi^{JPC}_{4q}\rangle$.}
    \label{tetradiagrams}
\end{figure}

Contributions to the Hamiltonian expectation value are summarized in Fig. \ref{tetradiagrams} and correspond to
4 self-energy, 6 scattering, 4
annihilation and 70 exchange terms each of which can be reduced to 12 dimensional integrals that are evaluated in momentum space. Because of the computationally intensive nature of this analysis, the
hyperfine interaction was not included.
Performing large-scale Monte Carlo calculations (typically 50 million samples), has conclusively determined that the molecular representation (i.e. meson-meson) produces the lightest mass state for a given $J^{PC}$. This is due to suppression of certain interactions (cancellation of  color factors) in the singlet-singlet molecular representation and also the presence of repulsive forces in the other,
more exotic,  color schemes.

\begin{figure}
   \includegraphics[width=.4\textwidth]{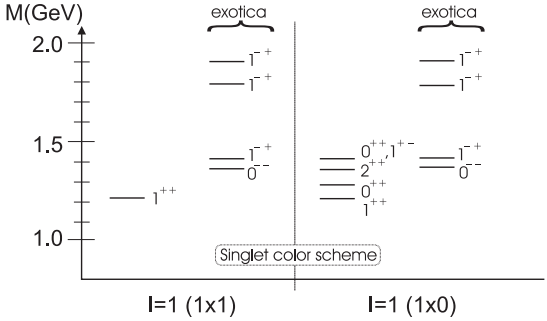}
   \caption{Tetraquark $I = $ 1 singlet scheme (molecule)  spectra.}
   \label{4qmolecule}
\end{figure}

\begin{figure}
   \includegraphics[width=.4\textwidth]{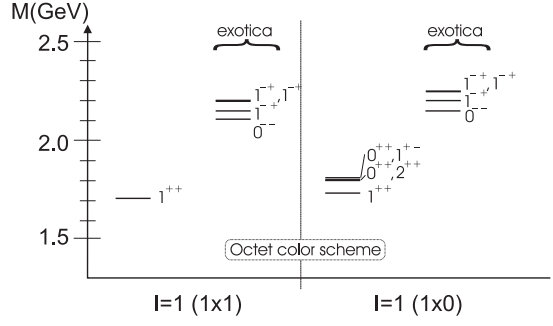}
   \caption{Tetraquark $I = $ 1 octet scheme spectra.}
   \label{4qatom}
\end{figure}

Using $m_u = m_d = 5$ MeV,  the predicted tetraquark ground state is the
non-exotic vector $1^{++}$ state in the molecular representation
with  mass around 1.2 GeV.   Figures \ref{4qmolecule} and
\ref{4qatom} depict the predicted tetraquark spectra for states having conventional
and exotic quantum numbers in both singlet and octet color representations.
Due to quark annihilation interactions ($q \bar{q} \rightarrow
g \rightarrow  q \bar{q}$) in the $I_{q \bar{q}} = 0$ channel
there are isospin splitting contributions, up to several hundred MeV, in the octet
but not singlet scheme as illustrated in the figures.  The annihilation interaction terms are repulsion, yielding octet states with $I =2$  lower than the $I = 1$ which are lower than the $I = 0$.  The molecular-type states are all  isospin degenerate and
the lightest exotic molecule  is the  $0^{--}$ with
mass 1.35 GeV.  Because of the isospin degeneracy, there will be several molecular-type
tetraquark states with the same $J^{PC}$ in the 1 to 2 GeV region.  Further, these states can be observed in different electric charge channels (different $I_z$) at about the same energy, which is
a useful experimental signature.
There are 3 orbital angular momentum and the lightest $1^{-+}$ exotics  have only 1 p-wave.  The
lightest $1^{-+}$ is predicted near 1.4 GeV which is close to the observed $\pi(1400)$, suggesting this state has a meson-meson type structure distinct from  states having quarks in  more exotic color representations (quark atoms).  Related, the computed tetraquark mass for $1^{-+}$ states with these more exotic color configurations is above 2 GeV.  This is consistent with  model predictions
\cite{gcl} for exotic hybrid meson ($q \bar{q} g$) $1^{-+}$ states also lying above 2 GeV due to repulsive color octet quark
interactions.  Finally,  for the same $J^{PC}$ states, including
the $1^{-+}$, the computed masses (not shown) in both the triplet  and the sextet diquark color representations are all heavier than in the singlet representation and  comparable to the octet
scheme results.

\section{Conclusions}
Summarizing, using the Coulomb gauge Hamiltonian model we have performed large-scale Monte Carlo calculations for $ggg$ glueballs, $q \bar q g$ hybrid mesons and $q \bar q q \bar q$ tetraquark
systems. Our oddball spectrum, along with lattice and other glueball model predictions,
clearly documents an odderon trajectory with slope similar to the pomeron but much lower intercept.
This explains why the odderon has not been
seen  in reactions with pomeron exchange.
If the odderon intercept is comparable to the $\omega$ value it may
be possible to observe it in reactions
where the pomeron is absent such as  pseudoscalar~\cite{eb1} or tensor
meson~\cite{eb2} electromagnetic production.
However if our prediction that the intercept is below 0.5
is correct, it is unlikely the odderon will be seen.  It is important therefore that
lattice calculations for the $5^{--}$ be performed  to confirm
this.

For the hybrid meson investigation, the Coulomb gauge Hamiltonian model
predicts that all $u \bar u g$ and $u \bar d g$ states have mass above 2 GeV.  In particular,
lattice and constituent model predictions for light, strange and charmed exotic $1^{-+}$ hybrid mesons are collectively in agreement.
This strongly suggest that the $\pi_1(1400)$ and $\pi_1(1600)$  are not  hybrid
mesons but have  an alternative structure as argued below.

Our tetraquark results clearly show that $[(q \bar q)_1 \otimes (q \bar q)_1]_1$ molecular states are lighter than the more exotic color octet $[(q \bar q)_8 \otimes (q \bar q)_8]_1$ atomic-like states.  Most significantly,
for the    $1^{-+}$, $I = 1$ channel our predicted lightest color octet state is above 2 GeV, while
the lightest singlet scheme state is around 1400 MeV, close to the $\pi(1400)$.
Combining these results with several model $1^{-+}$ hybrid meson predictions  indicates that the observed $\pi$ states are  not hybrids or exotic color configurations of 4 quarks but rather more conventional meson-meson type molecules.

Future work will address charmed ($c \bar{c} u \bar{u}$) and strange tetraquark ($c \bar{c} s \bar{s}$) systems to study the recently reported $X(3872)$ and $Y(4260)$ states.
Dynamical mixing of glueball, hybrid and conventional  meson and tetraquark states
will also be investigated along with
three-body forces
\cite{Szczepaniak:2006nx}.

\begin{acknowledgement}
The authors are very grateful to F. J. Llanes-Estrada and P. Bicudo for insightful discussions.
Work supported by   U. S. DOE grants
DE-FG02-97ER41048 and DE-FG02-03ER41260.
\end{acknowledgement}

%
%

\end{document}